\documentclass[showpacs,aps,prd]{revtex4}
\input epsf

\begin{document}
\title{Semileptonic decay $\Lambda_c\rightarrow\Lambda\ell^+\nu$
from QCD light-cone sum rules}
\author{Ming-Qiu Huang and Dao-Wei Wang}
\affiliation{Department of Physics, National University of Defense
Technology, Hunan 410073, China}
\date{\today}
\begin{abstract}
We present the study of the semileptonic decay
$\Lambda_c\rightarrow\Lambda\ell^+\nu$ by using the light-cone sum rule
approach. Distribution amplitudes(DAs) for the $\Lambda$ baryon are
discussed to the leading order conformal spin, and QCD sum rule estimate
for the corresponding parameters is presented. The form factors describing
the decay are calculated and used to predict the decay width and the decay
asymmetry parameter $\alpha$. With the inclusion of twist-3 contributions
the calculated decay width $\Gamma=(7.2\pm2.0)\times10^{-14}\mbox{GeV}$ as
well as asymmetry $\alpha=-(0.88\pm0.03)$ is found in good agreement with
the experimental data, while there are appreciable deviations from
experiment values when the higher twist contributions are included.
\end{abstract}
 \maketitle
\section{Introduction}
\label{sec1} The study of flavor changing decays of the $c$-quark
is always an active field in the heavy flavor physics, and the
most obvious reason lies on that those processes can provide
useful information on the various charm related
Cabibbo-Kobayashi-Maskawa(CKM) matrix elements which are the main
ingredients of the standard model(SM). Furthermore, the thorough
understanding of the SM in itself needs a comprehension of the
flavor changing dynamics. Unfortunately, such a comprehension is
difficult contemporarily, the fact is that form factors
characterizing those processes are not perturbative quantities and
whose determination must invoke some non-perturbative method. This
paper aims to give a preliminary determination of the form factors
of semileptonic $\Lambda_c\rightarrow\Lambda\ell^+\nu$ decay. In
the calculation we will use the method of QCD sum rules on the
light-cone \cite{BBK}, which in the past has been successfully
applied to various problems in heavy meson physics, see
\cite{lcsr} for a review.

The method of light-cone sum rules (LCSR) is a new development of
the standard technique of QCD sum rules \`a la SVZ sum rules
\cite{svzsum}, which comes as the remedy for the conventional
approach in which vacuum condensates carry no momentum
\cite{ball}. The main difference between SVZ sum rule and LCSR is
that the short-distance Wilson OPE in increasing dimension is
replaced by the light-cone expansion in terms of distribution
amplitudes of increasing twist, originally used in the description
of the hard exclusive process \cite{HEP}. In recent years there
have been many applications of LCSR to baryons.  The nucleon
electromagnetic form factors were studied for the first time in
\cite{nucleon1,nucleon2} and, more recently, in
\cite{nucleon3,nucleon4} for a further consideration. Several
nucleon related processes gave fruitful results within LCSR, the
weak decay $\Lambda_b\to p\ell\nu_\ell$ was considered in
\cite{lbtop} in both full QCD and HQET LCSR. The generalization to
the $N\gamma\Delta$ transition form factor was worked out in
\cite{ntodelta}.

In this paper we will adopt the LCSR approach to study the
exclusive semileptonic decay
$\Lambda_c\rightarrow\Lambda\ell^+\nu$. This transition had been
studied in the literature by several authors, employing flavor
symmetry or quark model or both in Refs.
\cite{Gavela,Perez,Singleton,Cheng,Migura}. There are also QCD sum
rule description of the form factors \cite{Dosch}, upon which the
total decay rate are obtained.

The paper is organized as follows: The relevant $\Lambda$ baryon
DAs are first discussed in Sec. \ref{sec2}. Following that Sec.
\ref{sec3} is devoted to the LCSRs for the semileptonic
$\Lambda_c\rightarrow\Lambda\ell^+\nu$ decay form factors. The
numerical analysis and our conclusion are presented in Sec.
\ref{sec4}.
\section{The $\Lambda$ baryon Distribution Amplitudes}\label{sec2}

Our discussion in this section for the $\Lambda$ baryon DAs parallels with
that for the nucleon \cite{hitwist}, so we only list the results following
from that procedure and for details it is recommended to consult the
original paper. The $\Lambda$ baryon DAs are defined through the matrix
element
\begin{eqnarray}
&&4\langle0|\epsilon_{ijk}u_\alpha^i(a_1 x)d_\beta^j(a_2
x)s_\gamma^k(a_3
x)|P\rangle=(\mathcal{A}_1+\frac{x^2M^2}{4}\mathcal{A}_1^M)(\rlap/P\gamma_5
C)_{\alpha\beta}\Lambda_{\gamma}+
\mathcal{A}_2M(\rlap/P\gamma_5C)_{\alpha\beta}(\rlap/x\Lambda)_{\gamma}
\nonumber\\&&{} + \mathcal{A}_3M(\gamma_\mu\gamma_5
C)_{\alpha\beta}(\gamma^\mu \Lambda)_{\gamma}+
\mathcal{A}_4M^2(\rlap/x\gamma_5C)_{\alpha\beta}\Lambda_{\gamma}+
\mathcal{A}_5M^2(\gamma_\mu\gamma_5
C)_{\alpha\beta}(i\sigma^{\mu\nu}x_\nu \Lambda)_{\gamma}+
\mathcal{A}_6M^3(\rlap/x\gamma_5C)_{\alpha\beta}(\rlap/x
\Lambda)_{\gamma},\label{das}
\end{eqnarray}
where the $\Lambda$ generically designates the spinor for the
$\Lambda$ baryon with momentum $P$. Only axial-vector DAs are
presented here, for those with other Lorentz structures do not
contribute in the final sum rules. The twist of those calligraphic
DAs is indefinite, but they can be related to the ones with
definite twist as
\begin{eqnarray}
&&\mathcal{A}_1=A_1, \hspace{2.4cm}2P\cdot
x\mathcal{A}_2=-A_1+A_2-A_3, \nonumber\\&& 2\mathcal{A}_3=A_3,
\hspace{2.2cm}4P\cdot x\mathcal{A}_4=-2A_1-A_3-A_4+2A_5,
\nonumber\\&& 4P\cdot x\mathcal{A}_5=A_3-A_4,\hspace{0.5cm}
(2P\cdot x)^2\mathcal{A}_6=A_1-A_2+A_3+A_4-A_5+A_6.
\end{eqnarray}
The twist of $A_i$ is given in Tab. \ref{twist}.
\begin{table}[htb]
\begin{tabular}{|c|c|c|c|}\hline
 twist-3 &{twist-4}&twist-5&twist-6\\\hline
$A_1$&$A_2$,$A_3$ &$A_4$,$A_5$&$A_6$ \\\hline
\end{tabular}
\caption{The twist for $A_i$.} \label{twist}
\end{table}
Each distribution amplitudes $F=A_i$ can be represented as Fourier
integral over the longitudinal momentum fractions $x_1$, $x_2$,
$x_3$ carried by the quarks inside the baryon with $\Sigma_i
x_i=1$,
\begin{equation}
F(a_iP\cdot x)=\int \mathcal{D}x e^{-ip\cdot
x\Sigma_ix_ia_i}F(x_i)\;.
\end{equation}
The integration measure is defined as
\begin{equation}
\int\mathcal{D}x=\int_0^1dx_1dx_2dx_3\delta(x_1+x_2+x_3-1).
\end{equation}
As elucidated in \cite{Braun99}, those distribution amplitudes are scale
dependent and can be expanded into orthogonal functions with increasing
conformal spin. To the leading conformal spin, or s-wave, accuracy the
expansion reads \cite{hitwist}
\begin{eqnarray}
A_1(x_i,\mu)&=&-\,120x_1x_2x_3\phi_3^0(\mu),\nonumber\\
A_2(x_i,\mu)&=&-\,24x_1x_2\phi_4^0(\mu),\nonumber\\
A_3(x_i,\mu)&=&-\,12x_3(1-x_3)\psi_4^0(\mu)
,\nonumber\\
A_4(x_i,\mu)&=&-\,3(1-x_3)\psi_5^0(\mu),\nonumber\\
A_5(x_i,\mu)&=&-\,6x_3\phi_5^0(\mu)\nonumber\\
A_6(x_i,\mu)&=&-\,2\phi_6^0(\mu), \label{da-a}
\end{eqnarray}
where the constraint $A(x_1,x_2,x_3)=A(x_2,x_1,x_3)$ arising from
the condition that the $\Lambda$ baryon has isospin $0$ has been
used in the derivation. All the $6$ parameters involved in Eq.
(\ref{da-a}) can be expressed in terms of $2$ independent matrix
elements of  local operators. Those parameters are
\begin{eqnarray}
\phi_3^0 = \phi_6^0 = -f_\Lambda,  \qquad \phi_4^0 = \phi_5^0 =
-\frac{1}{2} \left(\lambda_1 + f_\Lambda\right), \qquad
\psi_4^0 = \psi_5^0 = -\frac{1}{2}\left(\lambda_1-f_\Lambda
\right) \,. \nonumber
\end{eqnarray}
The normalization of $A_1$ at the origin defines the nucleon
coupling constant $f_\Lambda$,
\begin{equation}\label{fl}
\langle 0\mid \epsilon_{ijk}[u^i(0)C\gamma_5\rlap/ zd^j(0)]\rlap/
z s^k(0)\mid P\rangle=f_\Lambda z\cdot P\rlap/ z \Lambda(P).
\end{equation}
The remaining parameter $\lambda_1$ is defined by the matrix
element
\begin{equation}\label{l1}
\langle 0\mid \epsilon_{ijk}[u^i(0)C\gamma_5\gamma_\mu
d^j(0)]\gamma^\mu s^k(0)\mid P\rangle=\lambda_1 M \Lambda(P).
\end{equation}
The twist of the order $O(x^2)$ correction starts from twist five, which is
apparent in the definition (\ref{das}). Due to the numerically small
contribution it gives in the previous applications of light-cone QCD sum
rules \cite{lbtop}, we do not consider it in the following analysis.

\section{$\Lambda_c\rightarrow\Lambda\ell^+\nu$ decay form factors from light-cone Sum Rules}
\label{sec3}


Complying with the standard philosophy in the sum rule analysis,
we consider the following correlation function
\begin{equation}
\label{zt} z^\nu T_\nu(P,q)=iz^\nu\int d^4xe^{iq\cdot x}\langle
0\mid T\{j_{\Lambda_c}(0) j_\nu(x)\} \mid P\rangle,
\end{equation}
where $j_{\Lambda_c}=\epsilon_{ijk}(u^i C\gamma_5\rlap/z d^j)\rlap/z c^k$
is the current interpolating the $\Lambda_c$ baryon state, $j_ \nu=\bar
c\gamma_\nu(1-\gamma_5)s$ is the weak current, $C$ is the charge
conjugation matrix, and $i$, $j$, $k$ denote the color indices. The
auxiliary light-cone vector $z$ is introduced to project out the main
contribution on the light-cone. The interpolating current used here is not
the unique one, as exemplified in case studies \cite{current} for the
applications of QCD sum rules, and there can be other choices. The coupling
constant of the interpolating current to the vacuum can thus be defined as
\begin{equation}
\langle 0\mid j_{\Lambda_c} \mid
\Lambda_c(P')\rangle=f_{\Lambda_c}z\cdot P'\rlap/z\,\Lambda_c(P'),
\label{flc}
\end{equation}
where $\Lambda_c(P')$ and $P'$ is the $\Lambda_c$ baryon spinor
and four-momentum, respectively. Form factors are given in the
usual way
\begin{eqnarray}
\langle\Lambda_c(P-q)\mid j_\nu \mid \Lambda(P) \rangle&=&\bar
\Lambda_c(P-q)\left[f_1\gamma_\nu-i\frac{f_2}{M_{\Lambda_c}}\sigma_{\nu\mu}q^\mu
\right.\nonumber\\&-&\left.\left(g_1\gamma_\nu+i\frac{g_2}{M_{\Lambda_c}}
\sigma_{\nu\mu}q^\mu\right)\gamma_5\right]\Lambda(P),
\end{eqnarray}
in which $M_{\Lambda_c}$ is the $\Lambda_c$ mass, $\Lambda(P)$
denotes the $\Lambda$ spinor, satisfying
$\rlap/P\Lambda(P)=M\Lambda(P)$, where $M$ is the $\Lambda$ mass
and $P$ its four-momentum. Those form factors give no contribution
in the case of massless final leptons are omitted here.

Giving those definitions, the hadronic representation of the
correlation function (\ref{zt}) can be written as
\begin{equation}\label{ztha}
z^\nu T_\nu=\frac{2f_{\Lambda_c}}{M_{\Lambda_c}^2-P'^2}(z\cdot
P')^2\left[f_1\rlap/z+f_2\frac{\rlap/z\rlap/q}{M_{\Lambda_c}}-
\left(g_1\rlap/z-g_2\frac{\rlap/z\rlap/q}{M_{\Lambda_c}}\right)\gamma_5\right]\Lambda(P)+\cdots,
\end{equation}
where $P'=P-q$ and the dots stand for the higher resonances and
continuum. While on the theoretical side, at large Euclidean
momenta $P'^2$ and $q^2$ the correlation function (\ref{zt}) can
be calculated perturbatively and the result, in the leading order
of $\alpha_s$, is
\begin{equation}\label{ztth}
z^\nu
T_\nu=-2(C\gamma_5\rlap/z)_{\alpha\beta}\rlap/z(1-\gamma_5)_\mu\int
d^4x\int\frac{d^4k}{(2\pi)^4}\frac{z\cdot
k}{k^2-m_c^2}\;e^{i(k+q)\cdot x}\;\langle 0\mid
\epsilon_{ijk}u^i_\alpha(0)d^j_\beta(0) s^k_\mu(x)\mid P\rangle\;,
\end{equation}
where $m_c$ is the $c$-quark mass.Substituting (\ref{das})
into Eq. (\ref{ztth}) we obtain,
\begin{eqnarray}
z^\nu T_\nu&=&-2(z\cdot P)^2\left[\int dx_3\;
\frac{x_3B_0(x_3)}{k^2-m_c^2}
+M^2\int dx_3\;\frac{x_3^2B_1(x_3)}{(k^2-m_c^2)^2}+2M^4\int
dx_3\;\frac{x_3^3B_2(x_3)}{(k^2-m_c^2)^3}\right]
\rlap/z(1-\gamma_5)\Lambda(P) \nonumber\\&&+2(z\cdot
P)^2\left[M\int dx_3\;\frac{x_3B_3(x_3)}{(k^2-m_c^2)^2}+2M^3\int
dx_3\;\frac{x_3^2B_2(x_3)}{(k^2-m_c^2)^3}\right] \rlap/z\rlap/q
(1+\gamma_5)\Lambda(P)+\cdots,\label{ztres}
\end{eqnarray}
where $k=x_3P-q$ and the ellipses stand for contributions that are
nonleading in the infinite momentum frame kinematics
$P\rightarrow\infty$, $q\sim \mbox{const.}$, $z\sim 1/P$. The
functions $B_i$ are defined by
\begin{eqnarray}
B_0&=&\int_0^{1-x_3}dx_1A_1(x_1,1-x_1-x_3,x_3),\nonumber\\
B_1&=&-2\tilde{A_1}+\tilde{A_2}-\tilde{A_3}-\tilde{A_4}
+\tilde{A_5},\nonumber\\
B_2&=&\tilde{\tilde{A_1}}-\tilde{\tilde{A_2}}
+\tilde{\tilde{A_3}}+\tilde{\tilde{A_4}}-\tilde{\tilde{A_5}}
+\tilde{\tilde{A_6}},\nonumber\\
B_3&=&-\tilde{A_1}+\tilde{A_2}-\tilde{A_3}.
\end{eqnarray}
The DAs with tildes are defined via integration as follows
\begin{eqnarray}
\tilde{A}(x_3)&=&\int_1^{x_3}dx'_3\int_0^{1-x'_3}dx_1A(x_1,1-x_1-x'_3,x'_3),\nonumber\\
\tilde{\tilde{A}}(x_3)&=&\int_1^{x_3}dx'_3\int_1^{x'_3}dx''_3\int_0^{1-x''_3}dx_1
A(x_1,1-x_1-x''_3,x''_3).
\end{eqnarray}
The origin of those functions can be traced back to the partial
integration adopted to eliminate the factor $1/P\cdot x$ which
appears in the distribution amplitudes. When the next-to-leading
order conformal expansion is considered, the surface terms
completely sum to zero. The term $B_0$ corresponds to the leading
twist contribution. The form factors $f_2$ and $g_2$ in
(\ref{ztres}) are characterized by the higher twist contributions.

Equating (\ref{ztha}) and (\ref{ztres}), adopting the quark-hadron
duality assumption and employing a Borel improvement of $P'^2$ on
both sides lead us to the desired sum rules for the form factors
$f_1$ and $f_2$,
\begin{eqnarray}\label{sr-f1}
&&-f_{\Lambda_c} f_1
e^{-M_{\Lambda_c}^2/M_B^2}=-\int_{x_0}^1dx_2\;e^{-s'/M_B^2}\left[
B_0+\frac{M^2}{M_B^2}\left(-B_1(x_3)
+\frac{M^2}{M_B^2}B_2(x_3)\right)\right]\nonumber\\&&+\frac{M^2x_0^2e^{-s_0/M_B^2}}{m_c^2+Q^2+x_0^2
M^2}\left[B_1(x_0)
-\frac{M^2}{M_B^2}x_0B_2(x_0)\right]+\frac{M^2e^{-s_0/M_B^2}x_0^2}
{m_c^2+Q^2+x_0^2M^2} \frac{d}{dx_0}\left(\frac{M^2 x_0^2B_2(x_0)
}{m_c^2+Q^2+x_0^2M^2} \right) ,
\end{eqnarray}
and
\begin{eqnarray}\label{sr-f2}
&&\frac{f_{\Lambda_c} f_2}{M_{\Lambda_c}M}
e^{-M_{\Lambda_c}^2/M_B^2}=\frac{1}{M_B^2}\int_{x_0}^1\frac{dx_3}{x_3}\;e^{-s'/M_B^2}\left(
B_3(x_3)-\frac{M^2}{M_B^2}B_2(x_3)\right)+\frac{x_0e^{-s_0/M_B^2}}{m_c^2+Q^2+x_0^2M^2}
\nonumber\\&&\left(B_3(x_0)-\frac{M^2}{M_B^2}x_0B_2(x_0)\right)+
\frac{M^2e^{-s_0/M_B^2}x_0^2}{m_c^2+Q^2+x_0^2M^2}
\frac{d}{dx_0}\left(\frac{x_0B_2(x_0)}{m_c^2+Q^2+x_0^2M^2}\right),
\end{eqnarray}
where
\begin{equation}
s'=(1-x)M^2+\frac{m_c^2+(1-x)Q^2}{x},
\end{equation}
and $x_0$ is the positive solution of the quadratic equation for
$s'=s_0$:
\begin{equation}
2M^2x_0=\sqrt{(Q^2+s_0-M^2)^2+4M^2(Q^2+m_c^2)}-(Q^2+s_0-M^2).
\end{equation}
As the sum rules for the form factors $g_1$ and $g_2$ are
identical with those for the $f_1$ and $f_2$, $f_1=g_1$ and
$f_2=g_2$, we will only discuss the results for $f_1$ and $f_2$ in
the following sections.

\section{numerical analysis and the conclusion}\label{sec4}

\subsection{Values for $f_\Lambda$ and $\lambda_1$}

Parameters $f_\Lambda$, $\lambda_1$ appear in the conformal
expansion of the DAs, so we have to determine their values before
proceeding to analyze the LCSRs. According to their definitions,
we consider correlation functions
\begin{equation}
\label{cf-corr} \Pi(q^2)=i\int d^4xe^{iq\cdot x}\langle 0\mid
T\{J_i(x) \bar J_j(0)\} \mid 0\rangle.
\end{equation}
where $J_i's$ are given in (\ref{fl}) and (\ref{l1}). Following
the standard QCD sum rule philosophy, an estimate for
$f_\Lambda$,$\lambda_1$ and their relative sign is straightforward
\begin{eqnarray}\label{f1-sr}
(4\pi)^4f_\Lambda^2e^{-M^2/M_B^2}&=&\frac{2}{5}\int_{{m_s^2}}^{s_0}
s(1-x)^5e^{-s/M_B^2}ds-\frac{b}{3}\int_{{m_s^2}}^{s_0}
x(1-x)(1-2x)e^{-s/M_B^2}\frac{ds}{s},
\end{eqnarray}
\begin{eqnarray}\label{f2-sr}
4(2\pi)^4\lambda_1^2M^2e^{-M^2/M_B^2}&=&\frac{1}{2}\int_{{m_s^2}}^{s_0}
s^2\left[(1-x^2)(1-8x+x^2)-12x^2\ln\;x\right]e^{-s/M_B^2}ds\nonumber\\
&+&\frac{b}{12}\int_{{m_s^2}}^{s_0}
(1-x)^2e^{-s/M_B^2}ds-\frac{4}{3}\;a^2e^{-M^2/M_B^2},
\end{eqnarray}
\begin{eqnarray}\label{sign-sr}
4(2\pi)^4f_\Lambda\lambda_1^*Me^{-M^2/M_B^2}&=&\frac{m_s}{6}\int_{{m_s^2}}^{s_0}
s\left[(1-x)(3+13x-5x^2+x^3)+12x\ln\;x\right]e^{-s/M_B^2}ds\nonumber\\
&+&\frac{b}{12}\int_{{m_s^2}}^{s_0}
(1-x)\left[1+\frac{1}{3}(1-x)(5-\frac{2}{x})\right]e^{-s/M_B^2}\frac{ds}{s},
\end{eqnarray}
where $x=m_s^2/s$ and $m_s$ is the $s$-quark mass. The sum rule for the
$f_\Lambda$ has been obtained before \cite{lbtop} where the corresponding
heavy quark limit is also derived. At the working window
$s_0\sim1.6^2\;\mbox{GeV}^2$ and $1<M_B^2<2\;\mbox{GeV}^2$ the numerical
value for the coupling constant reads
\begin{eqnarray}
f_\Lambda=6.1\mathcal{\Theta}10^{-3}\mbox{GeV}^2,\hspace{0.3cm}
\lambda_1=-1.2 \mathcal{\Theta}10^{-2} \mbox{GeV}^2.
\end{eqnarray}
The relative sign of $\lambda_1/f_\Lambda$ is obtained from the sum rule
(\ref{sign-sr}). In the numeric analysis, the standard values
$a=-(2\pi)^2\langle \bar qq \rangle=0.55\mbox{GeV}^3$, $b=(2\pi)^2\langle
\alpha_sG^2/\pi \rangle=0.47~\mbox{GeV}^4$ and $m_s=0.15\mbox{GeV}$ are
adopted. It should be noted that our value for $f_\Lambda$ here does
coincide with that obtained in \cite{Chernyak}.

The sum rule for the coupling constant of $\Lambda_c$ to vacuum is
similar to that for $f_\Lambda$, where the simple substitution
$m_s\rightarrow m_c$ should be made, and the numerical value is
$f_{\Lambda_c}=(6.4\pm0.7)\mathcal{\Theta} 10^{-3}\mbox{GeV}^2$,
taken from the interval $1<M_B^2<2\mbox{GeV}^2$ with
$s_0\sim10\mbox{GeV}^2$.

\subsection{Analysis of the LCSRs}
In the numerical analysis for the form factors, the charm quark
mass is taken to be $m_c=1.41~\mbox{GeV}$ \cite{mc}, and the other
relevant parameters, $\Lambda_c$ and $\Lambda$ baryon masses and
the value of $|V_{cs}|$, can be found in \cite{PDG2004}. We start
with analysis of the twist-3 sum rules, in which only twist-3 DA
is kept. Substitute the above given parameters into the LCSRs and
vary the continuum threshold within the range
$s_0=7-9~\mbox{GeV}^2$, we find there exist an acceptable
stability in the range $M_B^2=5-7~\mbox{GeV}^2$ for the Borel
parameter. The $M_B^2$ and the $q^2$ dependence for the
\begin{figure}[b]
\centerline{\epsfysize=6truecm \epsfbox{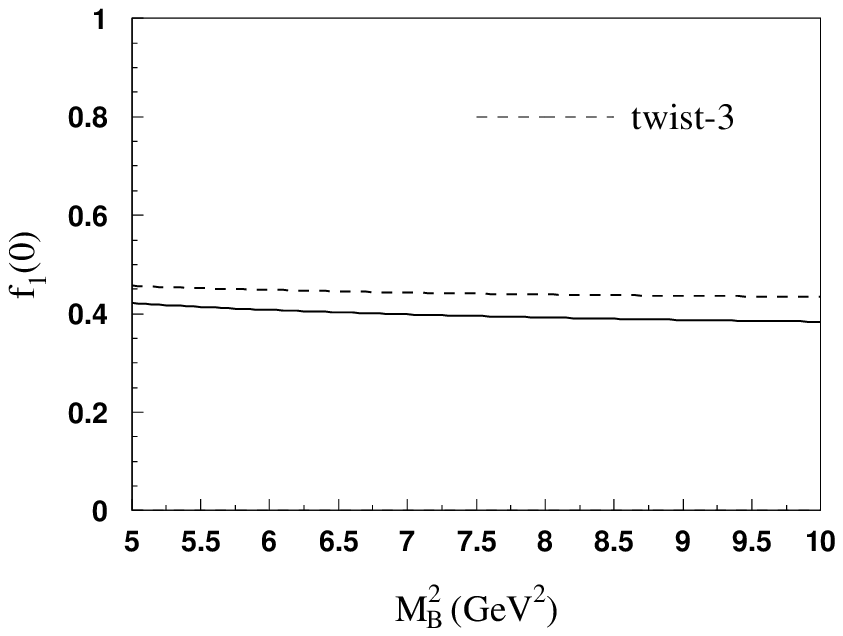}
\epsfysize=6truecm \epsfbox{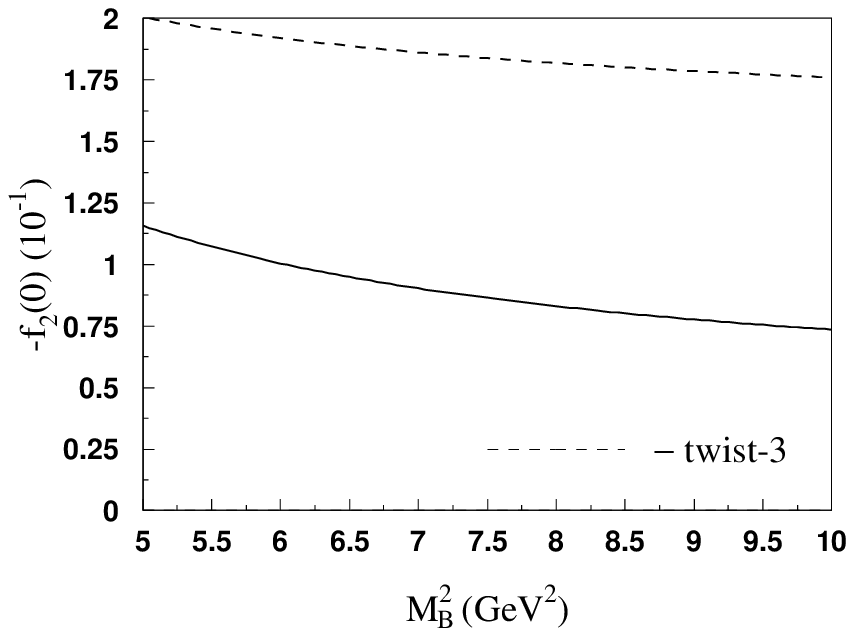}} \caption{The dependence on
$M_B^2$ of the LCSRs for the form factors $f_1$ and $f_2$ at
$q^2=0$. The continuum threshold is $s_0=8\mbox{GeV}^2$ for the
twist-3 result and $s_0=10\mbox{GeV}^2$ for the up to twist-6
one.} \label{fig:1}
\end{figure}
corresponding form factors are shown in Figs. 1 and 2,
\begin{figure}[t]
\centerline{\epsfysize=6truecm \epsfbox{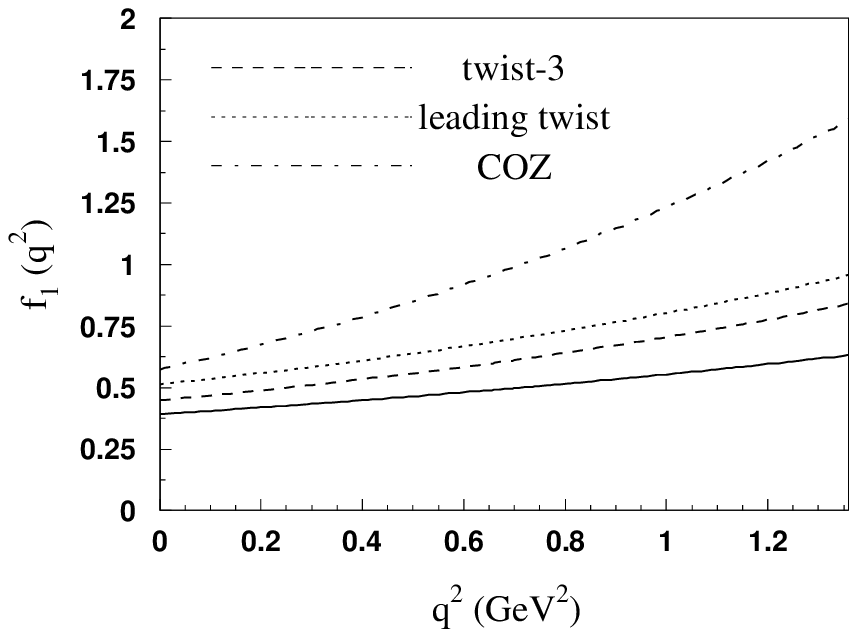}
  \epsfysize=6truecm \epsfbox{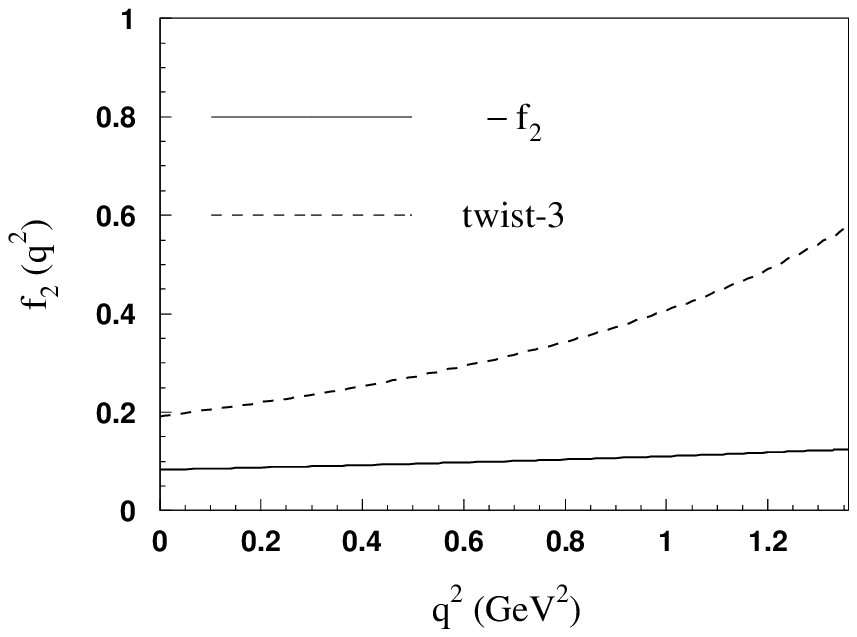}}
\caption{The dependence on $q^2$ of the LCSRs for the form factors
$f_1$ and $f_2$. The ``COZ" denotes the result obtained from the
COZ DAs. The continuum threshold and the Borel parameter are
$s_0=8\mbox{GeV}^2$, $M_B^2=6\mbox{GeV}^2$ and
$s_0=10\mbox{GeV}^2$, $M_B^2=8\mbox{GeV}^2$ for the twist-3 and
the up to twist-6 results. For the two leading twist results, they
are chosen to be the same with those for the twist-3 sum rule.}
\label{fig:2}
\end{figure}
respectively. Also given in Fig. 2 are two leading twist results,
corresponding to only retain $B_0$ in the sum rules. It is
apparent that in that approximation, only $f_1$ and $f_2$ survive.
Apart from the leading twist DA we discuss in (\ref{da-a}), there
still exists another form from Chernyak, Ogloblin and
Zhitnitsky\cite{Chernyak}:
\begin{equation}
A^{COZ}_1(x_i)=-21\varphi_{as}[0.52(x_1^2+x_2^2)+0.34x_3^2-2.05x_1x_2-0.48x_3(x_1+x_2)],
\label{coz}
\end{equation}
where $\varphi_{as}=120x_1x_2x_3$ is the asymptotic DA. The
corresponding result is also illustrated for contrast.

For the up to twist-6 sum rules, the stability is agreeable within the
range $s_0=9-11~\mbox{GeV}^2$ and $M_B^2=7-9~\mbox{GeV}^2$. In that working
region, the twist-3 contribution to $f_1$ is the dominant one, amounting to
over $90\%$. However, the case is different for $f_2$: the main
contribution comes from the twist-4 DAs and its magnitude is approximately
$\sim 1.5$ of the twist-3 one in the whole dynamical region, but with a
different sign. On account of the relatively small momentum transfer, the
asymptotic behavior of DAs may not be fulfilled and we need incorporate
higher conformal spin in the expansion for them. Furthermore, QCD sum rule
tends to overestimate the higher conformal spin expansion parameters
\cite{nonlocal}, and the corresponding parameter will enter in the
coefficients of the higher conformal spin expansion, which is well known as
the Wandzura-Wilczek type contribution. So at the current stage, the
strategy to stay with twist-3 sum rules seems to be a good choice. The
$M_B^2$ and the $q^2$ dependence for the corresponding form factors are
also shown in Figs. 1 and 2.

Both the form factors in the whole kinematical region,
$0<q^2<(M_{\Lambda_c}-M)^2$, can be fitted well by the dipole
formula
\begin{equation}
f_i(q^2)=\frac{f_i(0)}{a_2(q^2/M_{\Lambda_c}^2)^2+a_1q^2/M_{\Lambda_c}^2+1},
\label{dipole}
\end{equation}
and the uncertainties are negligible. Below in Table \ref{di-fit}
we give those coefficients for two sets of parameters:
$M_B^2=6~\mbox{GeV}^2$, $s_0=8~\mbox{GeV}^2$ for the twist-3 sum
rule and $M_B^2=8~\mbox{GeV}^2$, $s_0=10~\mbox{GeV}^2$ for the up
to twist-6 one.
\begin{table}[htb]
\begin{tabular}{|@{\hspace{1ex}}c|@{\hspace{1ex}}*{3}{r@{.}l
}@{\hspace{1ex}}|@{\hspace{1ex}}*{3}{r@{.}l}@{\hspace{1ex}}|}\hline
&\multicolumn{6}{c|@{\hspace{1ex}}}{twist-3}&\multicolumn{6}{c|}{up to twist-6}\\
\hline &\multicolumn{2}{c}{$a_2$}&\multicolumn{2}{c}{$a_1$}
&\multicolumn{2}{c|@{\hspace{1ex}}}{$f_i(0)$}&\multicolumn{2}{c}{$a_2$}
&\multicolumn{2}{c}{$a_1$}&\multicolumn{2}{c|}
{$f_i(0)$}\\
$f_1$&$1$&$595$&$-2$&$203$&$0$&$449$&$0$&$993$&$-1$&$712$&$0$&$392$ \\
$f_2$&$2$&$992$&$-3$&$329$&$0$&$193$&$0$&$238$&$-1$&$339$&$-0$&$083$ \\
\hline
\end{tabular}
\caption{The dipole fit for the form factors $f_1$ and $f_2$ with
$M_B^2=6~\mbox{GeV}^2$, $s_0=8~\mbox{GeV}^2$ for the twist-3
results and $M_B^2=8~\mbox{GeV}^2$, $s_0=10~\mbox{GeV}^2$ for the
up to twist-6 sum rules, (\ref{sr-f1}) and (\ref{sr-f2}).}
\label{di-fit}
\end{table}

Using the obtained form factors, we can calculate the differential
decay rate and the total decay width for the decay
$\Lambda_c\rightarrow\Lambda\ell^+\nu$. The differential decay
rate is shown in Fig. 3. If only the twist-3 amplitude $A_1$ is
retained, we have for the total decay width
$\Gamma=(7.2\pm2.0)\times10^{-14}\mbox{GeV}$, which agrees well
with the data given by the Particle Data Group \cite{PDG2004}.
This result is also in agreement with the QCD sum rule predictions
made in \cite{Dosch}. As for the decay asymmetry parameter
$\alpha$ defined in \cite{Korner}, we obtain
$\alpha=-0.88\pm0.03$, which corresponds to the ratio at zero
momentum transfer $f_2(0)/f_1(0)=0.44\pm0.05$. That value lies
very close to the recent experimental data from CLEO
\cite{Hinson}. Note that the errors quoted above reflect the
uncertainty due to the Borel parameter $M_B$ and the continuum
$s_0$. The uncertainty due to the variation of the other QCD
parameters is not included, which may reach $5\%$ or more.

\begin{figure}[t]
\centerline{\epsfysize=7.5truecm \epsfbox{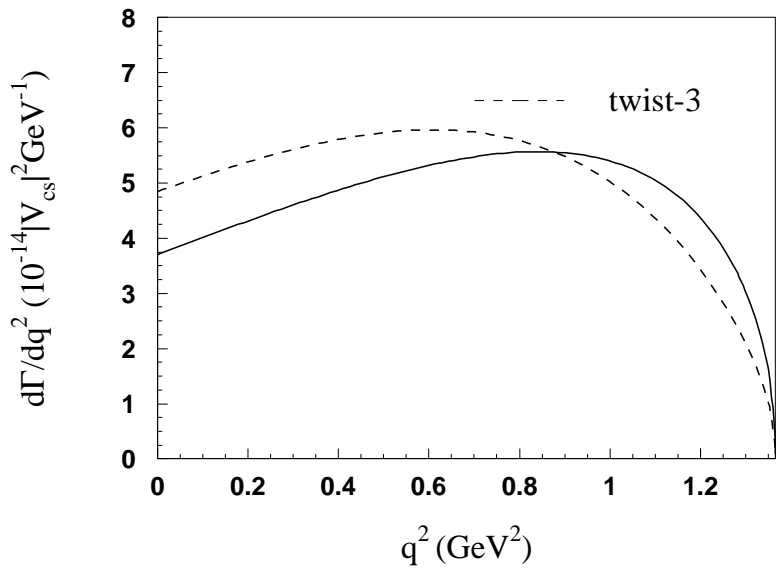}} \caption{Differential
decay rate for $\Lambda_c\rightarrow\Lambda\ell^+\nu$. The parameters
follow those in Fig. 2.} \label{fig:3}
\end{figure}

For a comparison, the total decay width computed from the up to
twist-6 form factors is
$\Gamma=(6.8\pm2.0)\times10^{-14}\mbox{GeV}$, whose agreement with
the experimental value is good, too. However, for the asymmetry
parameter $\alpha$, we get $\alpha=-(0.54\pm0.02)$, which lies
above the particle data group's average \cite{PDG2004} and still
greater than the latest experimental measurement \cite{Hinson}.
The ratio of the two form factors at zero momentum transfer of
that asymmetry is $f_2(0)/f_1(0)=-0.22\pm0.03$.  This phenomenon,
i.e., the twist-3 result agrees better with the experiments, may
be attributed to our incomplete inclusion of the higher conformal
spin components in the expansion for the DAs.

To summarize, we have given a preliminary investigation on the
semileptonic decay $\Lambda_c\rightarrow \Lambda\ell^+\nu$ using
the LCSR method. Sum rules for the form factors are derived and
used to calculate the decay width and the asymmetry parameter. The
decay width agrees well with the experimental data both for the
twist-3 and the up to twist-6 sum rules, while the asymmetry's
agreement is not so good: the twist-3 DA alone can account for the
experimental value but the result including higher twist
contributions is not so good. This is partly due to the interplay
of our incomplete inclusion of the higher conformal spin
contributions and the QCD sum rule over-estimated value for
$\lambda_1$.


\acknowledgments M.Q.H. would like to thank the Abdus Salam ICTP
for warm hospitality. This work was supported in part by the
National Natural Science Foundation of China.


\end{document}